# Optimization Approach for Detecting the Critical Data on a Database

Prashanth Alluvada
Department of Electrical Engineering
University of Arkansas, Fayetteville
Email: palluvad@uark.edu

**Abstract**

Through purposeful introduction of malicious transactions (tracking transactions) into randomly select nodes of a (database) graph, soiled and clean segments are identified. Soiled and clean measures corresponding those segments are then computed. These measures are used to repose the problem of critical database elements detection as an optimization problem over the graph. This method is universally applicable over a large class of graphs (including directed, weighted, disconnected, cyclic) that occur in several contexts of databases. A generalization argument is presented which extends the critical data problem to abstract settings.

## 1. Introduction

When transactions are made with the database, the data elements get modified, added or deleted. Evidently, further transactions which access the modified data elements carry the modified elements through the database. If the initial transaction is malicious of nature then all transactions which use the modified data (now corrupt) carry the malice through the database and corrupt other data on the database. Thereby the malice spreads through the entire database and corrupts large parts of the data it holds. It is therefore very important to identify and protect the "critical" or "sensitive data" elements which may be "most accessed" or "most sought after" data elements.

Since mission critical data protection is prime objective of IT industry, a novel data monitoring system is devised ([1]) which gives access to IT personnel to identify anomalous behavior during database transactions. For fixed probabilities of accessing the delay-sensitive objects, it is shown that partitioning the set of disks is always better than striping in all of the disks ([2]). A method to measure the extent of information leak and loss of important information is proposed [3], which develops a measure for the adversaries of the system. It is shown that through finite checks, adversaries of specific families may be related to a probabilistic information model.

In this article, we demonstrate that through purposeful introduction of malicious transactions the problem of critical data detection is reposed as an optimization problem over graphs. We further pose the generalized critical data detection problem and develop methods for introducing deterministic and linguistic constraints for the critical data detection.

## 2. Problem Formulation

<u>Definition 2.1:</u> the database is a weighted directed graph G(V,E). V is the set of vertices, W is the set of weights corresponding to the edges E.

For instance, through suitable graphing conventions the log file of the transactions may be pictured as a graph. The vertices or nodes may represent the data elements and the edges, the transactions. Weights on the edges may represent a normalized frequency of the corresponding transaction. They are assumed positive.

Definition 2.2: Pick a random subset of the nodes V of G(V,E) and purposefully introduce malicious (or tracking) transaction into those nodes. The transaction propagates along the edges of G(V,E). The subset of the graph affected by the transactions is the soiled segment. The unaffected part is the clean segment. The ratio of the sum of weights of the soiled segment to the sum of all weights on the graph is the measure of the soiled segment called soiled measure (S).

In the definitions, we consider the most general situation: the graph may possess cycles and may be disconnected. The cycles may be counted once while computing the weight contribution from their edges. The weights are assumed positive. Clearly, the soiled measure S is a function of the graph connectivity. Furthermore, every subset of the nodes has a corresponding soiled measure and graph's connectivity determines that measure. We note that when purposeful malice is planted into all nodes of G then trivially the entire graph is soiled while on the other hand a single node which soils all of G is the most critical node. Between these extremes, a smallest possible subset of the nodes which produces the largest possible soiled measure may be defined as the most critical subset of the G. Consequently, we have the following definition

Definition 2.3: A subset V' of V of n nodes is said to be critical if it is a global maximum of the following cost function:

$$F := S^2 + \beta(1 - \frac{n}{N})^2 \qquad (2.1)$$

where S is the soiled measure due to the n nodes into which purposeful malice is introduced, N is the total nodes on the graph.

It is clear from the Definition 2.3 that the two opposing factors of nodes (as small as possible) against the soiled measure due to those nodes (as large as possible) cause the cost function F to attain a maximum at the critical subset of the G. A global maximum of F is the "break-even" combination of the largest soiled measure against the smallest subset of the nodes the graph may admit. The second term at Eqn. (2.1) indicates that larger subsets of the nodes receive less credit of being critical subsets. $\beta$ is a suitably chosen tuning constant or a function of n and may be set according to the scaling requirement of the optimization problem. The function F at Eqn. (2.1) may be optimized through any of the random search methods (such as simulated annealing or genetic algorithm, [5]). The cost function Eqn. (2.1) provides a clear definition for the idea of the critical subset of a graph. Clearly, if G(V,E) admits a single critical node (which soils the entire graph) then Eqn. (2.1) attains a maximum at that node (S=1 and the second term is the largest possible).

Remarks: 1) Depending upon the data constraints, ease of computation etc, a suitable definition for soiled measure may be used. The main objective is to assign a suitable measure to the soiled segment. 2) The cost function at Eqn. (2.1) is one of the several possible functions which attains maximum at the break even point. For instance, for any K>0, KF is maximum at point P if and only if F is maximum at the same point. The choice at Eqn. (2.1) is sufficient for the problem. 3) Choice of $\beta$ is problem specific, may be positive or otherwise and should preserve the interesting sets targeted in the problem.

Example: Consider the graph shown at Fig. 2.1. Applying definition 2.3, the critical subset of the graph is the global maximum of the cost function at Eqn. (2.1). Choosing unity for $\beta$, the function we seek to optimize in order to recover the critical subsets is

$$S^2 + (1 - \frac{n}{N})^2 \tag{2.2}$$

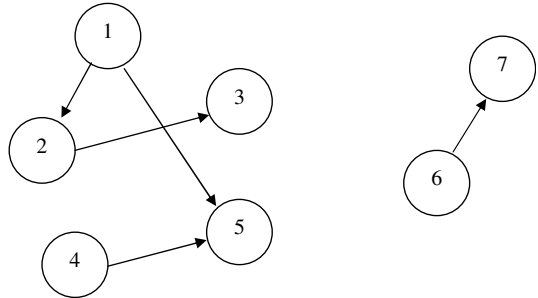

Fig. 2.1 A sample graph whose critical subset is sought

We assume that the weights of the graph are equal to one. Computations with cost function Eqn. (2.2) are shown at Table 2.1. The soiled measure shown on the last column indicates that the nodes 1,4,6 is the most critical subset (we note that they belong in two different connected sets).

The cost function Eqn. (2.2) may be enhanced to include that less nodes is more preferred. In this example, such function may be tailored by choosing $\beta(n) = 2/n$. Repeating the computation of Table 2.1 with the new cost function at Eqn. (2.3), the new computation is at Table 2.2

$$S^2 + \frac{2}{n}(1 - \frac{n}{N})^2 \tag{2.3}$$

| TABLE 2.1: Cost Function with beta=1 | | | | | |
|---|---|---|---|---|---|
| Serial Number | Purposeful malice into Nodes (n) | Soiled measure (S) | Clean measure | $E=S^2+(1-n/N)^2$ | E |
| 1 | 1,6 | 6/7 | 1/7 | $36/49+(1-2/7)^2$ | 1.2449 |
| 2 | 3,5 | 2/7 | 5/7 | $4/49+(1-2/7)^2$ | 0.5918 |
| 3 | 2,4,7 | 5/7 | 2/7 | $25/49+(1-3/7)^2$ | 0.8367 |
| 4 | 2,6 | 4/7 | 3/7 | $16/49+(1-2/7)^2$ | 0.8367 |
| 5 | 1,4,6 | 7/7 | 0 | $1+(1-3/7)^2$ | 1.3265 |

| TABLE 2.2: Cost Function with beta=2/n | | | | | |
|---|---|---|---|---|---|
| Serial Number | Purposeful malice into Nodes (n) | Soiled measure (S) | Clean measure | $E=S^2+2/n*(1-n/N)^2$ | E |
| 1 | 1,6 | 6/7 | 1/7 | $36/49+2/2*(1-2/7)^2$ | 1.2449 |
| 2 | 3,5 | 2/7 | 5/7 | $4/49+2/2*(1-2/7)^2$ | 0.5918 |
| 3 | 2,4,7 | 5/7 | 2/7 | $25/49+2/3*(1-3/7)^2$ | 0.7279 |
| 4 | 2,6 | 4/7 | 3/7 | $16/49+2/2*(1-2/7)^2$ | 0.8367 |
| 5 | 1,4,6 | 7/7 | 0 | $1+2/3*(1-3/7)^2$ | 1.2177 |

The new cost function (Eqn. (2.3)) has determined that the nodes 1,6 is the most critical subset. The switch from the result of Table 2.1 is evident.

## 3. Extending the Definition of Critical Data

Taking cues from the arguments developed in the previous section and the Eqn. (2.1), the idea of critical data is extended to the abstract setup of the generalized critical data.

Definition 3.1 A global maximum of the following cost function

$$\alpha S^2 + \beta(\gamma - \delta \frac{n}{N})^2 + \sum_k \varepsilon_k C_k \qquad (3.1)$$

is the generalized critical subset of G(V,E). The $\alpha, \beta, \gamma, \delta$ and $\varepsilon$ are suitably chosen coefficient functions (of n as well as other parameters of the problem) and the $C_k$ are constraints which extend the definitions of critical data sets. The coefficient functions map from suitable subsets of natural numbers to Real numbers (Complex numbers).

Remarks: 1) through coefficient functions $\alpha, \beta$, the function types for the soiled measure and the nodes (decreasing function for increasing nodes) may be independently controlled. Essentially a desired function may be incorporated for the soiled measure and the nodes terms by setting the $\alpha$ and $\beta$ accordingly. 2) within admissible limits of the problem, the definition of the critical sets may be varied through appropriate choices of $\alpha, \beta, \gamma, \delta$ and $\varepsilon$.

The constraints $C_k$ augment the cost function through the Lagrange multipliers $\varepsilon_k$. The coefficients $\alpha, \beta, \gamma, \delta$ and $\varepsilon$ may be chosen according to the design requirements of the problem. The constraints $C_k$ may be abstract or even linguistic. The critical data problem is generalized through the introduction of the coefficient functions and the constraints which may be deterministic or even linguistic. Through optimization of Eqn. (3.1) we pose the "generalized critical data detection" problem. For the linguistic case, the augmentation is performed through the AND operator and the cost function is implemented as $\alpha S^2 + \beta(\gamma - \delta \frac{n}{N})^2$ AND (CONSTRAINTS).

This function (for the deterministic or the linguistic case) may be optimized through any of the random search methods (such as simulated annealing or genetic algorithm [5]). Fuzzy constraints may also be included through the composition rules of Fuzzy logic and implemented through the AND operator ([4]). We demonstrate an application of the linguistic constraint with the generalized cost function by continuing with the example and seeking critical subsets "within the same connected component."

Example: In continuation of the example at Fig. 2.1, we impose a linguistic constraint that the critical subset belongs in the same connected component of the graph. This linguistic constraint is implemented through the AND operator thus:
Using Eqn. (3.1), set $\alpha(n) = 1$, $\beta(n) = 1$, $\gamma(n) = 1$, $\delta(n) = 1$, $\varepsilon_1 = 1$ and define the cost function

$$E = S^2 + (1 - \frac{n}{N})^2 \text{ AND } C_1 \qquad (3.2)$$

$C_1$, the constraint, is implemented in the following fashion. If the nodes $\{n_1, n_2,...,n_k\}$ are randomly picked then
$C_1=1$ iff $\{n_1, n_2,...,n_k\}$ all belong in the same connected component of the graph, else 0.
$$\qquad (3.3)$$
Further the cost function is defined via $C_1$ as:
$$E = S^2 + (1 - \frac{n}{N})^2 \text{ iff } C_1=1; \text{ undefined if } C_1=0. \qquad (3.4)$$

Table 3.1shows the calculation for the cost function defined through Eqns. (3.3), (3.4). E is not defined at the serial numbers 1,2,4,5 because by definition Eqn. (3.4), cost function is undefined

wherever $C_1=0$. Consequently, the calculation over connected components has determined that nodes 1,4 (serial number 6) is the most critical within any connected component.

| TABLE 3.1: Cost Function together with Linguistic Constraint | | | | | | |
|---|---|---|---|---|---|---|
| Serial Number | Purposeful malice into Nodes (n) | $C_1$ | Soiled measure (S) | Clean measure | $E=S^2+(1-n/N)^2$ | E |
| 1 | 1,6 | 0 | 6/7 | 1/7 | - | - |
| 2 | 3,5 | 1 | 2/7 | 5/7 | $4/49+(1-2/7)^2$ | 0.5918 |
| 3 | 2,4,7 | 0 | 5/7 | 2/7 | - | - |
| 4 | 2,6 | 0 | 4/7 | 3/7 | - | - |
| 5 | 1,4,6 | 0 | 7/7 | 0 | - | - |
| 6 | 1,4 | 1 | 5/7 | 2/7 | $25/49+(1-2/7)^2$ | 1.0204 |
| 7 | 6 | 1 | 2/7 | 5/7 | $4/49+(1-1/7)^2$ | 0.8163 |
| 8 | 2,4 | 1 | 4/7 | 3/7 | $16/49+(1-2/7)^2$ | 0.8367 |

**4. Advantages**

The cost function approach to critical data items detection offers several advantages: 1) globally optimal combination (most critical data) is obtained. This is the most critical subset the graph may admit. Theoretically, no better combination is possible 2) every possible graph topology is included. Be it weighted, directed, disconnected or their combination, simulated annealing (random search) includes all. 3) through the language of soiled and clean measures, most analyses of the database graph may be reposed conveniently and suitably in terms of cost functions. 4) new constraints may be easily and instantly incorporated into the problem by augmenting the cost function with the constraint (deterministic or linguistic) and performing the optimization over the augmented cost function. For the deterministic case, the Lagrange multiplier method may be used for function augmentation. 5) linguistic constraints may also be incorporated and linguistic variables may be used for implementing the linguistic constraints (as illustrated in the example, using the AND operator) 6) this approach provides flexibility (through various choices of the coefficient functions, Eqns (3.1), (3.2)) for the critical data detection problem of databases 7) designer's heuristics and experience with the database may be fruitfully encoded into the coefficient functions ($\alpha, \beta, \gamma, \delta$ and $\varepsilon$) for that database. 8) Cost functions provide a unified treatment to critical data detection problem over a large class of data sets.

**5. Conclusion and Future Work**

Through the language of clean and soiled measures, the notion of critical subset of the graph is posed as an optimization problem over the graph. Extending further, the generalized critical subset is defined as the global maximum of a generalized cost function, obtained by means of a function augmentation through Lagrange multipliers. Deterministic and linguistic constraints are used for augmenting the given cost function. The optimization approach for critical data detection is flexible and includes a wide choice for cost functions for varied data types encountered in practice.

The main idea introduced in this article is purposeful introduction of "tracking signals" through randomly selected nodes of the graph and associating a measure to the subsets of the graph the tracking signals have reached. Using this measure, it is demonstrated that important and interesting subsets of the graph may be characterized through a family of optimization problems. This approach may be employed in a general setting of the graphs and important subsets may be detected through the corresponding optimization problems. The tracking signals may be pre-designed to capture the suitable subsets of the graph. In future publications, author shall demonstrate the various applications and extensions of this method (through deterministic, linguistic and abstract types of constraints.)